\documentclass[a4,12pt]{article}
\newcommand{\be}{\begin{eqnarray}}
\newcommand{\ee}{\end{eqnarray}}
\makeatletter
\newcommand{\fslash}[2][0mu]{%
    \mathchoice
     {\fsl@sh\displaystyle{#1}{#2}}%
     {\fsl@sh\textstyle{#1}{#2}}%
     {\fsl@sh\scriptstyle{#1}{#2}}%
      {\fsl@sh\scriptscriptstyle{#1}{#2}}}
    \newcommand{\fsl@sh}[3]{%
    \m@th\ooalign{$\hfil#1\mkern#2/\hfil$\crcr$#1#3$}}
\def\lsim{\raise0.3ex\hbox{$<$\kern-0.75em\raise-1.1ex\hbox{$\sim$}}}
\def\gsim{\raise0.3ex\hbox{$>$\kern-0.75em\raise-1.1ex\hbox{$\sim$}}}
\makeatother
\usepackage{epsfig}
\usepackage{graphics}

\title{\bf Hadronic fluctuations at the \\
QCD phase transition}
  \author{ \\ S. Ejiri$^{1,2}$, F. Karsch$^{1,3}$ and
  K. Redlich$^{4,5}$
    \\ \\
    \small
$^1$ Fakult\"at f\"ur Physik, Universit\"at Bielefeld,
D-33615 Bielefeld, Germany\\
\small $^2$ Department of Physics, The University of Tokyo,
Tokyo 113-0033, Japan \\
\small $^3$ Physics Department, Brookhaven National Laboratory,
Upton, NY 11973, USA \\
\small $^4$ Institute of Theoretical Physics, University of
Wroclaw,
PL-50204 Wroclaw, Poland \\
\small $^5$ Physics Department, Theory Division, CERN, CH-1211 Geneva 23,
Switzerland
}
\begin{document}
\maketitle

\vspace{-12.5cm}

\hfill {\small BI-TP-2005-33}

\hfill {\small BNL-NT-05/28}

\hfill {\small CERN-PH-TH/2005-161}

\hfill {\small TKYNT-05/18 }

\vspace{12.5cm}

\centerline{Abstract} {\small We discuss the properties of
hadronic fluctuations, i.e. fluctuations of net quark and isospin
numbers as well as electric charge, in the vicinity of the QCD
transition in isospin-symmetric matter at vanishing quark chemical
potential. We analyse second- and fourth-order cumulants of these
fluctuations and argue that the ratio of quartic and quadratic
cumulants reflects the relevant degrees of freedom that carry the
quark number, isospin or charge, respectively.
In the hadronic phase we find that an enhancement of charge
fluctuations arises from contributions of doubly charged hadrons
to the thermodynamics. The rapid suppression of fluctuations seen
in the high-temperature phase suggests that in the QGP,  net quark
number and electric charge are predominantly carried by
quasi-particles, with the quantum numbers of quarks. }

\newpage
\section{Introduction}

In a medium of strongly interacting elementary particles,
described by equilibrium thermodynamics of Quantum Chromodynamics
(QCD), the transition between the low-temperature hadronic and
high-temperature quark--gluon phase occurs in a small temperature
interval and leads to a rapid change in entropy and energy
density. Although it is, most likely,  that this transition is not
a true phase transition, but rather a smooth but rapid crossover,
it is expected that this transition leads to large fluctuations of
energy and net quark number densities. In fact, lattice studies of
various hadronic susceptibilities, such as net quark number,
isospin or charge susceptibilities, show that quadratic
fluctuations of these quantum numbers increase rapidly in the
transition region \cite{Gottlieb,others}. These susceptibilities
continue to grow also above the transition temperature, $T_c$, and
come close to the ideal-gas value at temperatures $T\simeq 2T_c$.
Below $T_c$ bulk thermodynamics as well as the generic structure
of susceptibilities is found to be in good agreement with
properties of a hadronic resonance gas \cite{Tawfik}.

Recent lattice studies of 2-flavour QCD at non-zero quark
($\mu_q$) and isospin ($\mu_I$) chemical potential
[4--7] have sustain  that higher-order derivatives of the QCD
partition function (generalized susceptibilities), show even more
pronounced variations with temperature in the vicinity of the QCD
transition temperature. In particular, the fourth--order
derivatives with respect to $\mu_{q,I}$, or equivalently with
respect to the corresponding up and down quark chemical potentials
($\mu_u$, $\mu_d$), have pronounced peaks at the transition
temperature. As a consequence quark number fluctuations calculated
at non-zero $\mu_q$ increase with increasing values of $\mu_q$ and
develop a cusp along the transition line. This is in agreement
with expectations based on the (mean-field) analysis of quark
number fluctuations in sigma models \cite{Hatta} as well as NJL
models \cite{NJL}. These models share the relevant $O(4)$ symmetry
with 2-flavour QCD, and give some hint for the existence of a
second-order phase transition point in the QCD phase diagram at
some non-zero value of the quark chemical potential. In this paper
we will, however, concentrate on a discussion of hadronic
fluctuations at vanishing quark chemical potential. This is a
regime of the QCD phase diagram most relevant to current heavy-ion
experiments at RHIC as well as future experiments at the LHC.

Lattice calculations of bulk thermodynamics, e.g. of the equation
of state, as well as studies of generalized susceptibilities, show
significant deviations from the ideal-gas behaviour for a rather
large temperature range above $T_c$. In particular, the equation
of state shows sizeable deviations from the ideal-gas relation,
$\epsilon = 3p$, up to $T\simeq (3$--$4)T_c$. This is in line with
recent experimental findings that hint at the formation of a
strongly interacting medium in heavy-ion collisions at RHIC
energies \cite{rhic}.

It  has been recently argued that this strongly interacting medium
may, in fact, consist of a large set of heavy coloured bound
states, which could exist in a temperature interval $T_c\le T
\lsim ($1--$2)T_c$ \cite{shuryak,shuryak2}. The so-called sQGP
model suggests that these bound states can contribute as much as
20\% to the pressure in the high-temperature phase of QCD.
Such states clearly will also contribute to hadronic fluctuations.
We will discuss here to what extent fluctuations of such states are consistent
with lattice results on quark number and charge fluctuations in the QGP.

In the next section we will give a definition of the observables
we are going to analyse and present the lattice results we are
going to discuss in the following sections. In section 3 we
discuss the properties of quadratic and quartic fluctuations in
the vicinity of the QCD transition temperature. Section 4 is
devoted to a discussion of hadronic fluctuations in the low- and
high-temperature phases of QCD. Section 5 contains our
conclusions.

\section{Hadronic fluctuations at \boldmath $\mu_B=0$}

Quark number as well as charge and isospin fluctuations are
obtained from the grand canonical partition function of QCD
through appropriate combinations of derivatives with respect to
quark chemical potentials. As we will concentrate here on a
discussion of their properties at finite temperature and vanishing
quark chemical potential in 2-flavour QCD, we will use the QCD
partition function with non-vanishing up and down quark chemical
potentials only as a tool to generate the derivatives of interest.
After having done so the chemical potentials will be set to zero.

The pressure of a thermodynamic system is directly related to the
grand canonical partition function:
\begin{equation}
P(V,T,\mu_u, \mu_d) = \frac{T}{V} \ln Z(V,T,\mu_u, \mu_d) \quad ,
\label{lnZ}
\end{equation}
where $\mu_u$ and $\mu_d$ are the chemical potentials for up and down
quarks, respectively.

The fluctuations of quark number, isospin and  charge as well as
expectation values of higher-moments of these observables can be
obtained from generalized susceptibilities:
\cite{lgt3,Gavai,Gavai0}
 \be \chi^{n,m} = \frac{\partial^{n+m}
P}{\partial\mu_u^n\;\partial\mu_d^m} \quad . \label{general_chi}
\ee These susceptibilities, evaluated for vanishing quark chemical
potentials, define the coefficients of a Taylor series for the
pressure expanded simultaneously in terms of $\mu_u$ and $\mu_d$.
In the following we will consider only the mass-degenerate case
($m_u=m_d$) for which the susceptibilities are symmetric,
$\chi^{n,m}=\chi^{m,n}$. Recent lattice calculations have shown
that the fourth derivatives, $\chi^{4,0}$, as well as the
off-diagonal susceptibilities, $\chi^{2,2}$, become large in the
vicinity of the transition temperature \cite{lgt3}. Quartic
fluctuations thus seem to be particularly sensitive to details of
the transition to the high-temperature phase.

\begin{figure}[t]
\begin{minipage}[t]{65mm}
\includegraphics[width=6.3cm]{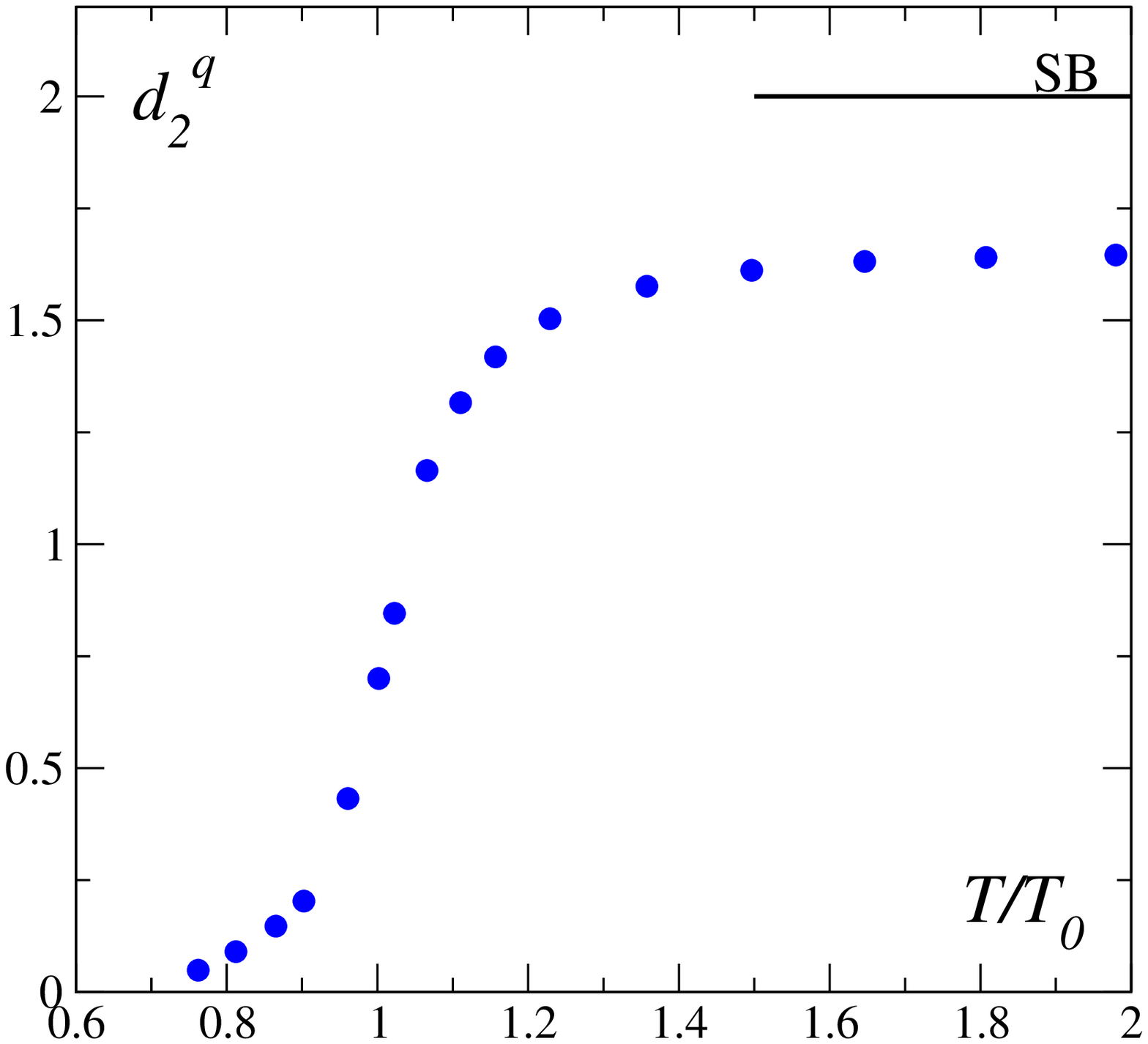}
\end{minipage}
\hskip 0.4cm
\begin{minipage}[t]{65mm}
\includegraphics[width=6.3cm]{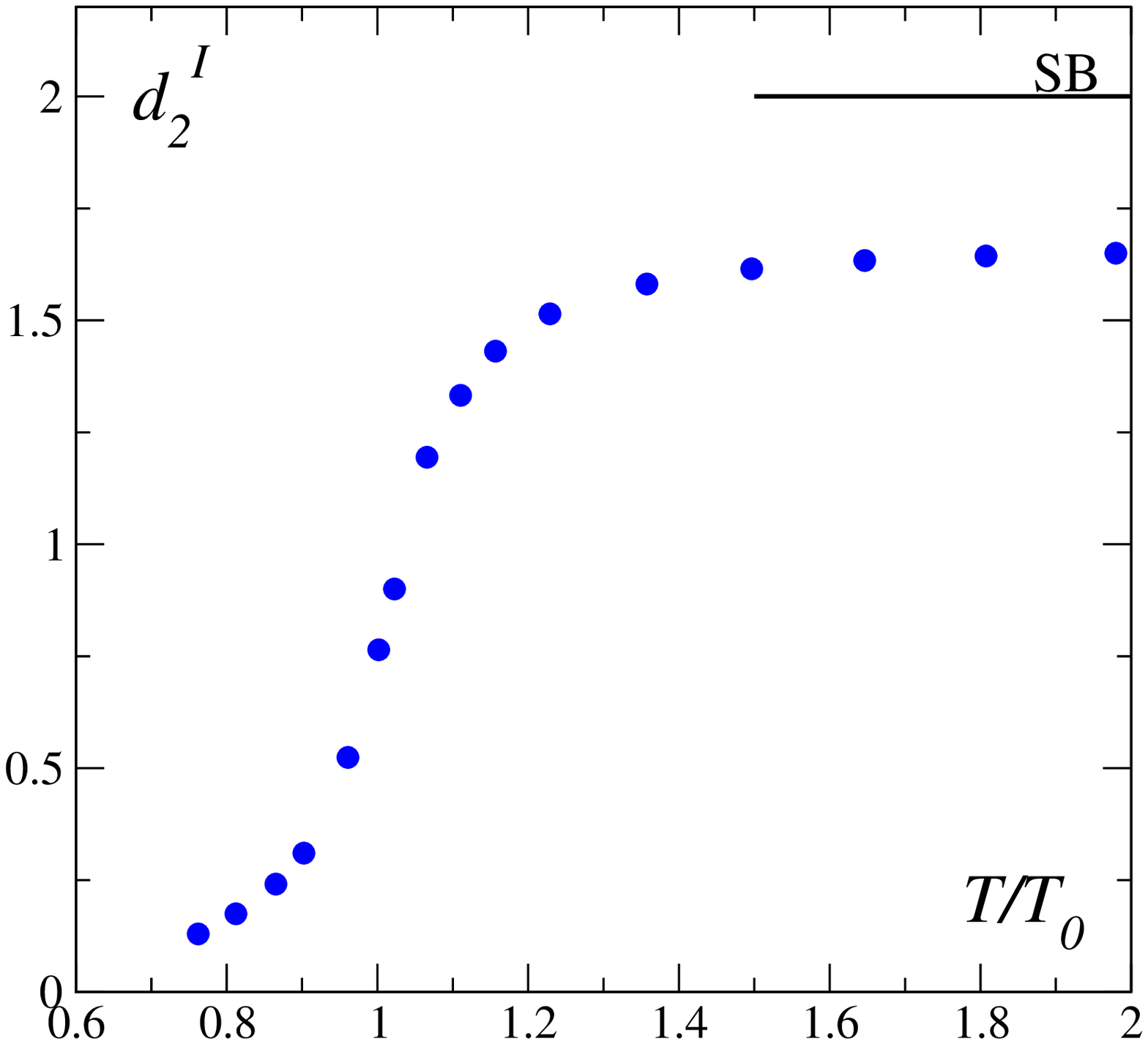}
\end{minipage}
\caption{\label{fig:d2}Quadratic fluctuations of the quark (left) and isospin
numbers.}
\end{figure}

Fluctuations of net quark ($N_q=\Delta N_u+\Delta N_d$), isospin
($N_I=\Delta N_u-\Delta N_d$) and charge ($Q=\frac{2}{3}\Delta N_u
- \frac{1}{3}\Delta N_d$) numbers, with $\Delta
N_{u,d}=N_{u,d}-\bar{N}_{u.d}$ denoting the difference of quarks
and antiquarks, are conveniently discussed in terms of quark and
isospin chemical potentials: \be \mu_q={{\mu_u+\mu_d}\over 2}~~,~
~ \mu_I={{\mu_u-\mu_d}\over 2} \label{eq2} \quad . \ee Charge
fluctuations can then be obtained from appropriate combinations of
derivatives with respect to $\mu_{u,d}$ or $\mu_{q,I}$,
respectively:
\begin{equation}
\frac{\partial}{\mu_Q} \equiv
\frac{2}{3}\frac{\partial}{\partial \mu_u}-
\frac{1}{3}\frac{\partial}{\partial \mu_d}=
\frac{1}{6}\frac{\partial}{\partial \mu_q}+
\frac{1}{2}\frac{\partial}{\partial \mu_I}
\quad .
\label{charge_deriv}
\end{equation}

The first two, non-vanishing derivatives with respect to $\mu_{q,I,Q}$
then yield
\begin{eqnarray}
d_2^x \equiv\frac{1}{VT^3}
\frac{\partial^2 \ln Z}{\partial (\mu_{x}/T)^2} &=& \frac{1}{VT^3}
\langle (\delta N_{x})^2 \rangle \quad ,\quad x=q,~I,~Q
\nonumber \\
d_4^x\equiv \frac{1}{VT^3}
\frac{\partial^4 \ln Z}{\partial (\mu_{x}/T)^4} &=& \frac{1}{VT^3}
\left( \langle (\delta N_{x})^4 \rangle - 3 \langle (\delta N_{x})^2
\rangle^2 \right)    \quad .
\label{fluctuations}
\end{eqnarray}
We note that the derivatives of the partition function with
respect to $\mu_{q}$ are directly related to the expansion
coefficients $c_k$ calculated in Refs.~\cite{lgt2,lgt3}.
Furthermore, $d_2^I$ is related to the expansion coefficient
$c_2^I$, while $d_2^Q$ can be obtained from a combination of $c_2$
and $c_2^I$, which  were  obtained also in these references.

\begin{figure}[t]
\begin{minipage}[t]{130mm}
\includegraphics[width=6.0cm]{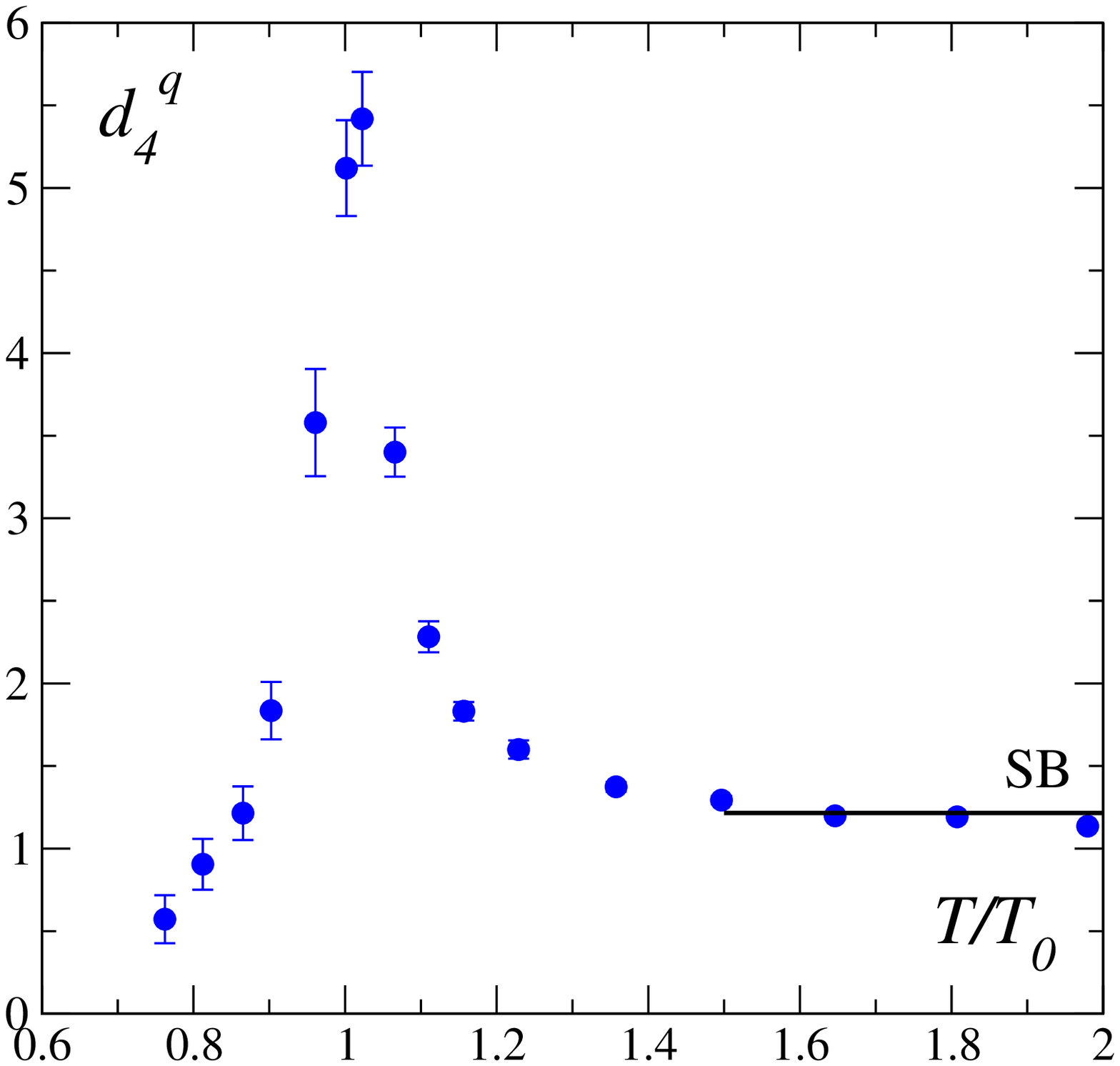}
\hskip 0.4cm
\includegraphics[width=6.0cm]{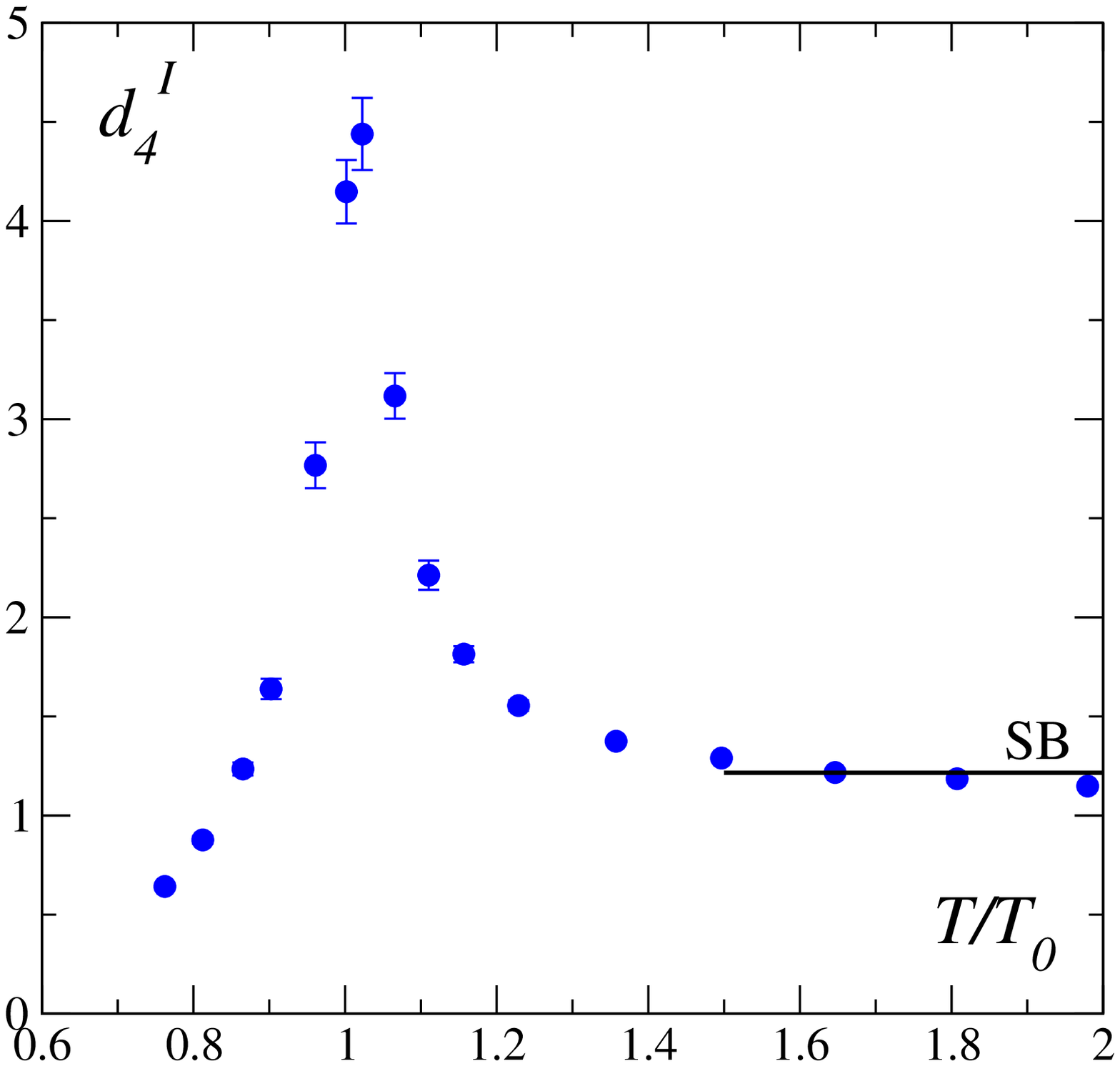}
\end{minipage}

\begin{minipage}[t]{60mm}
\includegraphics[width=6.0cm]{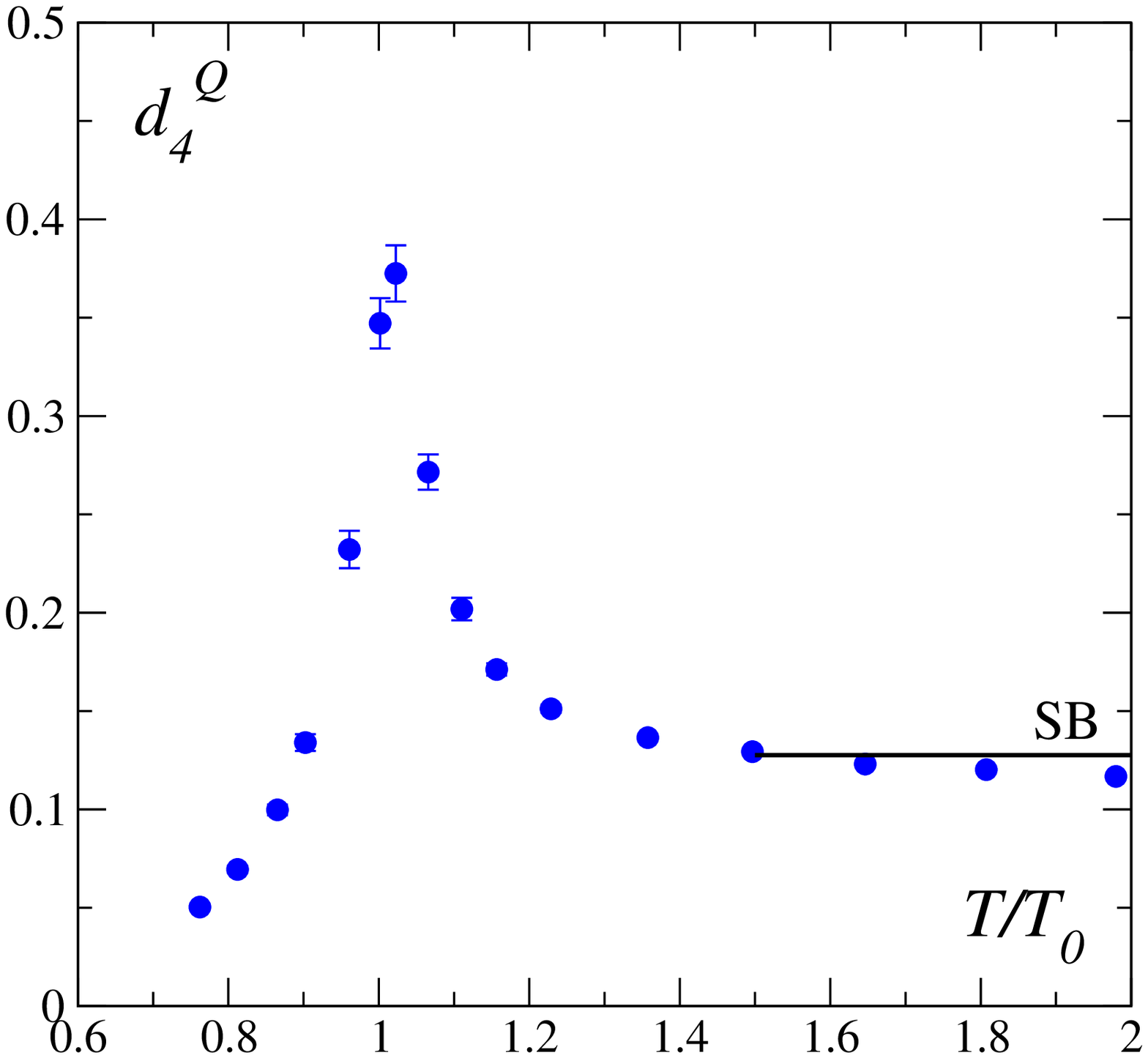}
\end{minipage}
\caption{\label{fig:d4}Fourth cumulant of the quark (top, left)
and isospin (top, right) number fluctuations as well as the
fourth cumulant of the charge fluctuations (bottom, left).}
\end{figure}

As is evident from Eq.~(\ref{charge_deriv}) charge fluctuations
can easily be expressed in terms of corresponding fluctuations in
quark and isospin (or net up and down quark) numbers. While the
quadratic charge fluctuations are entirely determined by the
corresponding quark number and isospin fluctuations, the quartic
charge fluctuations are also sensitive to correlations between
$\delta N_q$ and $\delta N_I$. They thus provide independent
information:
\begin{eqnarray}
d_2^Q &=& \frac{1}{6^2} \left( d_2^q + 9 d_2^I \right) \; ,
\nonumber \\[2mm]
d_4^Q &=& \frac{1}{6^4} \biggl( d_4^q +81 d_4^I
+ 54\left( \langle (\delta N_{q})^2 (\delta N_{I})^2 \rangle
-3  \langle \delta N_{q} \delta N_{I} \rangle^2 \right) \biggr) \; .
\label{charge}
\end{eqnarray}
In Fig.~\ref{fig:d2} we show the two independent second-order
cumulants $d_2^q$ and $d_2^I$. They are directly proportional to
the coefficients $c_2$ and $c_2^I$ calculated in
Refs.~\cite{lgt2,lgt3}. The three fourth-order
cumulants\footnote{These results are based on fourth-order
derivatives of the pressure calculated in Refs.~\cite{lgt2,lgt3}.}
are shown in Fig.~\ref{fig:d4}.

While the second-order cumulants show the well known continuous
increase towards the high-temperature ideal-gas value, the
fourth-order cumulants clearly have a pronounced peak at the
transition temperature and exceed the ideal-gas value by more than
a factor 4.  Here it is worthwhile to note that the lattice
results shown in Figs.~\ref{fig:d2} and \ref{fig:d4} are based on
calculations that were  performed with still rather heavy quarks,
corresponding to a pion mass of about $770$~MeV. Although we still
no have   satisfactory understanding of the universal behaviour of
the QCD transition in 2-flavour QCD, the studies of the quark mass
dependence of, for instance,   the chiral susceptibility [14--17]
suggest that the peaks
in susceptibilities increase with decreasing quark mass. We thus
also expect that the peaks found in fourth derivatives of the
pressure will increase with decreasing quark mass. In the next
section we briefly discuss the generic structure of second- and
fourth-order cumulants expected to be found in the chiral limit.

\section{Chiral symmetry restoration and
\\
hadronic fluctuations}

In the vicinity of the chiral transition temperature, and for
small values of the quark chemical potentials, the pressure
receives contributions from a regular and a singular part:
\begin{equation}
P(T,\mu_u, \mu_d) = P_r(T,\mu_u, \mu_d) + P_s(\bar{t},\bar{\mu}_u, \bar{\mu}_d)
\quad ,
\label{P_critical}
\end{equation}
where we have introduced in the singular part the reduced
temperature $\bar{t}=|T-T_c|/T_c$ and the dimensionless chemical
potentials $\bar{\mu}_{u,d} = \mu_{u,d}/T$. Deviations from
criticality are controlled by a generalized version of the reduced
temperature variable, which takes into account that the
temperature-like direction at non-vanishing quark chemical
potential in general will be a mixture of the couplings $\bar{t}$
and $\mu_q/T$. This generalized reduced temperature has to respect
the invariance of the QCD partition function under exchange of
particles and antiparticles ($\mu_q \rightarrow -\mu_q$):
\begin{eqnarray}
t&\equiv& \bar{t}+A (\bar{\mu}_u^2+\bar{\mu}_d^2)+
B \bar{\mu}_u \bar{\mu}_d \nonumber \\
~&=& \bar{t}+(2 A + B) \bar{\mu}_q^2+ (2 A - B) \bar{\mu}_I^2
\quad .
\label{reduced_T}
\end{eqnarray}
The singular part of the pressure may then be written as
\begin{eqnarray}
P_s(\bar{t},\bar{\mu}_u, \bar{\mu}_d) \sim b^{-1} f_s(b^{1/y_t} t)
\sim t^{2-\alpha}
\quad ,
\label{P_scaling}
\end{eqnarray}
where we have introduced an arbitrary scale parameter $b$ and
where $y_t=2-\alpha$ is the usual thermal critical exponent, which
is related to the critical exponent $\alpha$. Equation
 (\ref{P_scaling}) shows that the critical behaviour of hadronic
fluctuations $\mu_u=\mu_d=0$ is controlled by the critical
exponent of the specific heat, $\alpha$ \cite{Hatta}.
The contribution of the singular part to second- and fourth-order
cumulants of quark ($k\equiv q$) and isospin ($k\equiv I$) is then
given by
\begin{eqnarray}
\left( \partial^2 P_s/\partial \mu_k^2 \right)_{\mu_u=\mu_d=0} &\sim&
t^{1-\alpha} \; , \nonumber \\
\left( \partial^4 P_s/\partial \mu_k^4 \right)_{\mu_u=\mu_d=0} &\sim&
t^{-\alpha} \; .
\end{eqnarray}

For a conventional second-order phase transition with a critical
exponent $\alpha >0$ we would thus  expect that the fourth-order
cumulants diverge in the chiral limit. However, if the chiral
transition is indeed of second-order we expect it to belong to the
universality class of three-dimensional $O(4)$ symmetric spin
models. In this case the exponent $\alpha$ will be negative,
$\alpha \simeq -0.21$ \cite{Engels}. The contribution of the
singular part will then vanish at $t=0$ and the value of the
cumulants at $t=0$  will thus be determined by the regular part
alone. In the vicinity of $t=0$ the temperature dependence of the
second-order cumulant will then be dominated by the regular part.
This is in accordance with the rapid but smooth increase of second
cumulants across the transition temperature. The temperature
dependence of the fourth-order cumulant, however, will be
dominated by the singular part and as $\alpha$ is small the
temperature variation will result in a sharp cusp. This is in
accordance with the behaviour seen in Fig.~\ref{fig:d4}.

If the chiral transition in 2-flavour QCD is of second-order, we
thus expect the second- and fourth-order cumulants to stay finite
at the critical point. This, however, only holds for the critical
point at $\mu_q=0$, which we have discussed above. If the
transition in 2-flavour QCD at $\mu_q=0$ indeed belongs to the
$O(4)$ universality class, this transition point will be part of a
line of second-order phase transitions at non-zero quark chemical
potential. At a transition point with $\mu_q > 0$, one would then
 expect that already the second-order cumulant of quark number
fluctuations develops a cusp \cite{Hatta} and that the
fourth-order cumulant would diverge.

\section{Hadronic fluctuations in a thermal medium}

In the previous section we have discussed the qualitative
structure of second- and fourth-order cumulants in the vicinity of
the QCD transition temperature. Here we want to look in somewhat
more detail into the temperature dependence of these cumulants in
the low and high-temperature phases of QCD. In particular, we want
to argue that the ratios of fourth- and second-order cumulants of
quark number and charge fluctuations,
\begin{equation}
R_{4,2}^q = \frac{d_4^q}{d_2^q} \quad , \quad R_{4,2}^Q = \frac{d_4^Q}{d_2^Q}
\quad ,
\label{ratios}
\end{equation}
are sensitive observables that allow the  identification of  the
carriers of quark number and electric charge in a thermal medium.
We will first discuss this for a hadron gas at low-temperature and
then move on to a discussion of the relevant degrees of freedom in
the high-temperature phase. This discussion can easily be extended
to the case of isospin fluctuations, which, however, we will not
consider here any further.

\subsection{Fluctuations in hadronic matter}

We describe the low-temperature phase of QCD in terms of a
resonance gas with free particle dispersion relations for all
constituents \cite{Tawfik,model}. The resulting pressure, however,
also accounts for interactions between hadrons, to the extent that
the thermodynamics of an interacting system of elementary hadrons
can effectively be approximated by that mixture of ideal-gases of
stable particles and resonances \cite{model,hagedorn}. The
pressure $P$ is then expressed as a sum over all mesonic and
baryonic degrees of freedom:

\be P(T,\mu_u,\mu_d)=  P^M(T,\mu_u,\mu_d)+ P^B(T,\mu_u,\mu_d) \quad .
\label{eq1}
\ee

Using Eq.~(\ref{eq2}) to replace the up and down quark chemical
potentials by $\mu_q$ and $\mu_I$, the contribution of mesons,
baryons and their resonances belonging to fixed baryon and
isovector quantum numbers is obtained from Eq. (\ref{eq1}) in
compact form. With the exception of the pion, all hadrons are
heavy, with respect   to the temperatures of interest. For these
particles quantum statistics  plays no   role, and the Boltzmann
approximation of the partition functions is appropriate. This
yields, in the meson sector,

\be
{{P^M(T,\mu_I)}\over  {T^4}}=\sum_i F(T,m_i)
 \cosh {{2I_i^{(3)}\mu_I}\over T} \quad ,
\label{eq3}
 \ee
where  the sum is taken over all  non-strange mesons\footnote{In
order to be able to compare with lattice calculations performed
for 2-flavour QCD, we have neglected the contribution of strange
particles to the hadronic pressure.},   $I_i^{(3)}$ being  the
third component of the isospin of the species $i$. The function
$F(T, m_i)$ is the contribution of particles and the corresponding
antiparticles of mass $m_i$ with spin degeneracy factor $d_i$ to
the pressure at vanishing baryon and isospin chemical potential

\be F(T,m_i)=\frac{p_i}{T^4} =
{d_i\over {\pi^2}}\left( {m_i\over T}\right)^2 K_2\left( {{m_i}\over T}
\right) \quad . \label{eq5}
\ee

Neglecting the mass difference between  isospin partners, the
meson contribution arises from isospin singlet $G^{(1)}$
($I^{(3)}_i=0$) and triplet $G^{(3)}$ ($I^{(3)}_i=0,~\pm 1$)
mesons, respectively. Thus,

 \be
{{P^M(T,\mu_q,\mu_I)}\over  {T^4}} =G^{(1)}(T)+{1\over
3}G^{(3)}(T) [ 2\cosh ({{2\mu_I}\over T})+1 ] \quad . \label{eq3a}
 \ee

The contribution to the pressure arising from non-strange
baryons and their resonances is given by

 \be
{{P^B(T,\mu_q,\mu_I)}\over T^4} =\sum_i F(T,m_i)
 \cosh {{3\mu_q+2I_i^{(3)}\mu_I}\over T} \quad .
\label{eq4}
 \ee
\noindent The baryonic part of the pressure receives contributions
from isospin doublet $F^{(2)}$  ( $I^{(3)}_i=\pm 1/2$) and quartet
$F^{(4)}$ ($I^{(3)}_i=(\pm 1/2, \pm 3/2$)) baryons, respectively.
Again ignoring the mass differences between isospin partners, the
total baryonic contribution to the pressure is given by

 \be
{{P^B(T,\mu_q,\mu_I)}\over  {T^4}}&=& F^{(2)}(T) \cosh({{3\mu_q}\over T})
\cosh({{\mu_I}\over T})\nonumber \\&+&
 {1\over 2}F^{(4)}(T) \cosh({{3\mu_q}\over T})[\cosh({{\mu_I}\over
T}) +\cosh({{3\mu_I}\over T})] \quad .
\label{eq4a}
 \ee
We note that only the isospin quartet baryons contain doubly charged
hadrons, which is reflected in the appearance of a term proportional to
$\cosh (3\mu_I/T)$.

Within this resonance gas formulation we now can evaluate
cumulants of the quark number, isospin and charge in the confined
phase of QCD. In fact, by examining the  cumulants introduced in
section~2, we can find  the contributions of different hadronic
quantum number channels and can establish relations between  the
various cumulants. Within the Boltzmann approximation for the
hadron resonance gas, which we have introduced above, we find

\begin{eqnarray}
d_2^q(T) &=& 9\left( F^{(2)}(T)+F^{(4)}(T) \right)  \quad , \quad
\nonumber \\
d_4^q(T) &=& 81 \left( F^{(2)}(T)+F^{(4)}(T) \right)
\quad ,\label{eqq}
\end{eqnarray}


\begin{eqnarray}
d_2^Q(T) &=& {2\over 3} G^{(3)}(T)+{1\over 2}F^{(2)}(T)+{3\over 2} F^{(4)}(T)
\quad , \quad
\nonumber \\
d_4^Q(T) &=& {2\over 3}G^{(3)}(T)+{1\over 2}F^{(2)}(T)+{9\over 2} F^{(4)}(T)
\quad . \label{eqQ}
\end{eqnarray}
These equations are valid for isospin-symmetric systems at
vanishing baryon number density. The corresponding derivatives of
the contributions to the total pressure (Eqs.
(\ref{eq3}--\ref{eq4a})) were thus taken at $\mu_I=\mu_q=0$. The
resonance gas model leads to simple relations for the ratios of
cumulants of quark number and charge fluctuations. In particular,
we found
\begin{eqnarray}
R^q_{4,2} \equiv \frac{d^q_4}{d^q_2} &=&
\frac{\langle (\delta N_q)^4 \rangle}{\langle (\delta N_q)^2 \rangle}- 3
\langle (\delta N_q)^2 \rangle  =9 \quad ,
\label{R42q}
\end{eqnarray}
which is shown to agree well with the lattice results below $T_c$
\cite{lgt2,lgt3}.

The ratio $R^q_{4,2}$ is completely insensitive to details of the
hadron spectrum contributing to the resonance gas. In particular,
it reflects the fact that the quark number is carried only by
baryons,  i.e. the relevant degrees of freedom either carry net
quark number 0, in which case they do not contribute to
$R_{4,2}^q$, or 3, in which case they contribute to $R_{4,2}^q$
with a relative factor of $3^2$.

This is similar for the ratio of cumulants of charge fluctuations,
although we here have to take into account that hadrons with
charge $0,~1$ and $2$ can contribute to the thermodynamics. The
ratio $R_{4,2}^Q$  become sensitive to the relative abundance of
doubly charged hadrons:
\begin{eqnarray}
R^Q_{4,2} \equiv  \frac{d^Q_4}{d^Q_2} &=&
\frac{\langle (\delta Q)^4 \rangle}{\langle (\delta Q)^2 \rangle}- 3
\langle (\delta Q)^2 \rangle \nonumber \\
&=& {4 G^{(3)} + 3 F^{(2)} + 27 F^{(4)} \over
4 G^{(3)} + 3 F^{(2)} + 9 F^{(4)} }
\quad .
\label{R42Q}
\end{eqnarray}
At low-temperature, this ratio is dominated by charge fluctuations
in the pion sector, which contributes to $G^{(3)}$. The ratio will
thus approach unity at low-temperatures and  monotonically
increase  with temperature. This indicates that in the
low-temperature limit all charged degrees of freedom carry one
unit of charge. The ratio $R_{4,2}^Q$ does increase only from
contributions arising from isospin quartet baryons, which can
carry charge $Q=2$. In fact, denoting the contribution of all
hadrons carrying charge $Q=\pm n$ by $F_H^{Q=n}(T)$, one can
verify that Eq.~(\ref{R42Q}) can be rewritten as\footnote{This
becomes obvious when one reconstructs the origin of the algebraic
factors appearing in Eq.~(\ref{eqQ}). For instance, the factor
$3/2$ in front of $G^{(3)}$ indicates that, out of 3 triplet
states, 2 carry charge $|Q|= 1$ and the 3rd state does not
contribute as it carries charge $Q=0$. Similarly the factors $3/2$
and $9/2$ appearing in front of the quartet contribution $F^{(4)}$
arise from the fact that, out of the 4 states, 2 carry charge
$|Q|= 1$, 1 state carries charge $Q=2$ and the 4th state does not
contribute as it carries charge $Q=0$.}
\begin{equation}
R^Q_{4,2} =
\frac{F_H^{Q=1}(T) + 4 F_H^{Q=2}(T)}{F_H^{Q=1}(T) + 16 F_H^{Q=2}(T)} \quad .
\label{redone}
\end{equation}

\begin{figure}[t]
\begin{center}
\epsfig{file=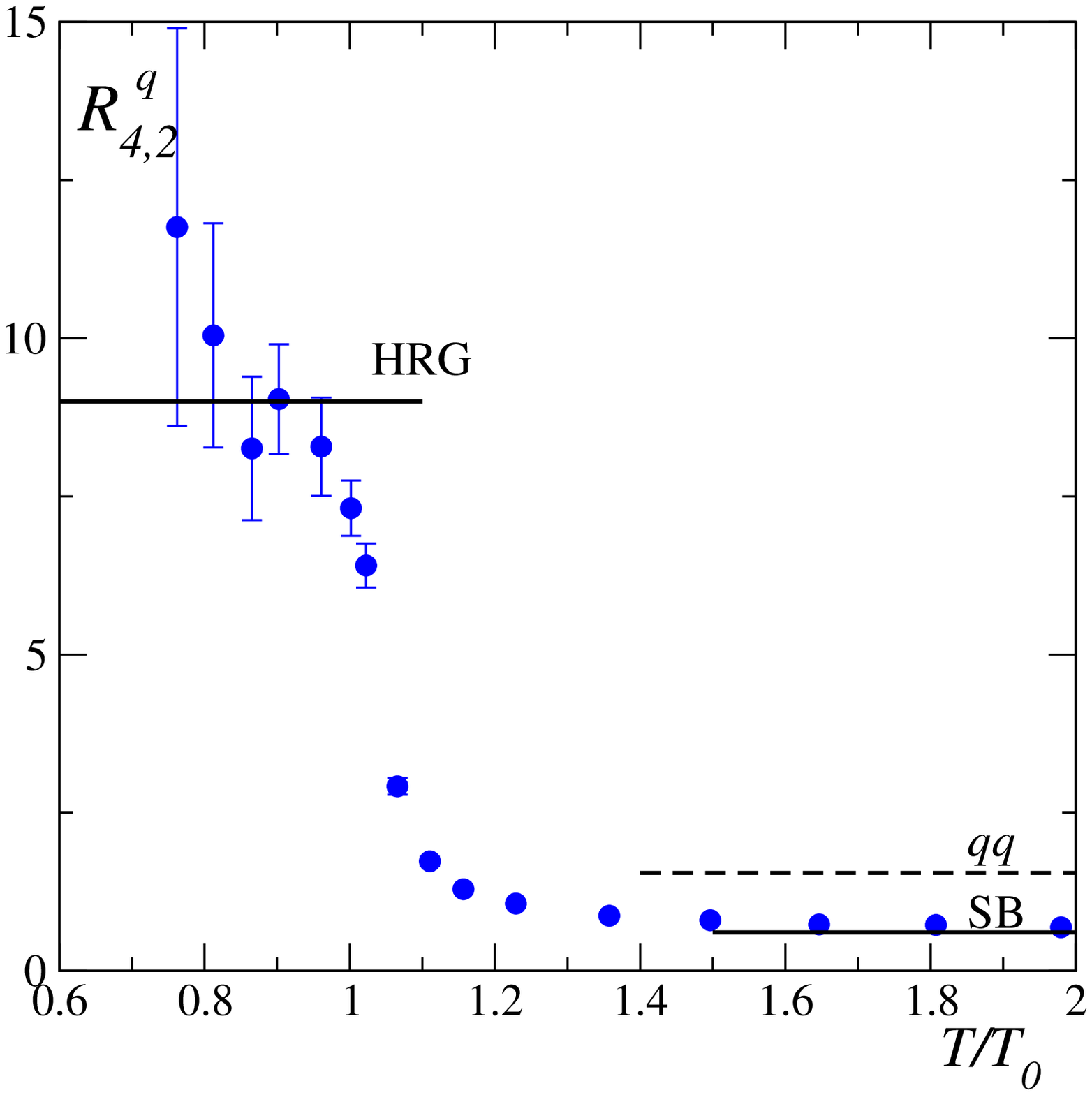,width=6.8cm}~\epsfig{file=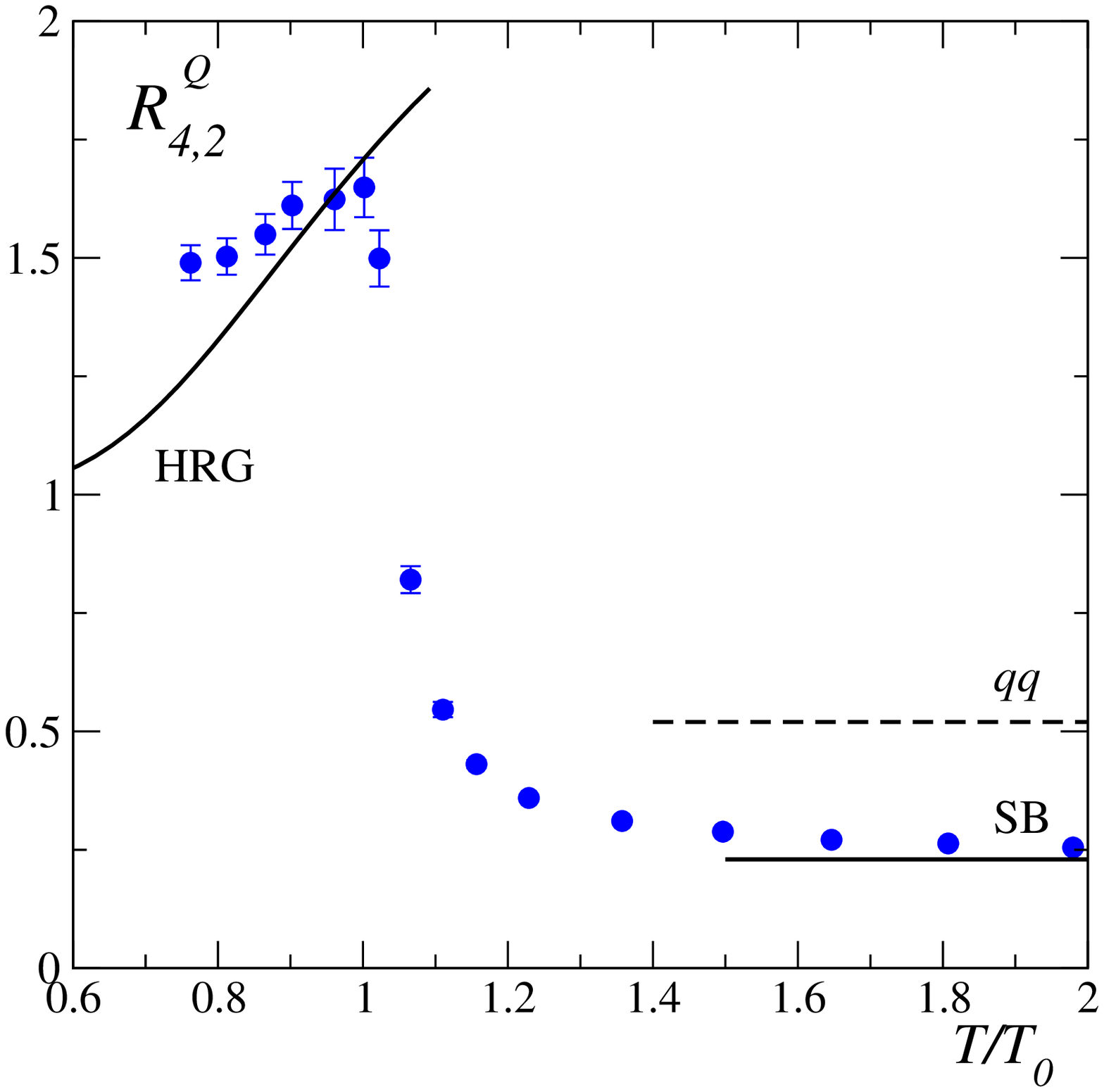,width=6.8cm}
\end{center}
\caption{\label{fig:charge} The ratios of fourth and second
cumulants of quark number (left) and charge (right) fluctuations.
At low-temperatures the ratios  $R_{4,2}^q$ and $R_{4,2}^Q$ are
compared to resonance gas model predictions. At high-temperature
the asymptotic ideal-gas value is indicated by a horizontal line.
The dashed line represents the estimate for the value of the
ratios at $T=1.4T_c$ in the presence of coloured $qq$ and $qg$
states asssuming that the $qq$ states contribute with half their
statistical weight as discussed in the text.}
\end{figure}

Results from lattice calculations for the ratios defined in
Eqs.~(\ref{R42q}) and (\ref{R42Q}), are shown in
Fig.~\ref{fig:charge}. In the low-temperature phase these ratios
clearly show the basic features expected from the analysis of a
hadron resonance gas. Above $T_c$, however, they rapidly drop and
strongly deviate from the resonance gas behaviour.
In fact, already at $T\simeq 1.5 T_c$ the ratios are close to the
corresponding ideal-gas values. In high-temperature perturbation
theory the leading ${\cal O}(g^2)$ corrections to quadratic and
quartic cumulants cancel in the corresponding ratios:
\begin{eqnarray}
R_{4,2}^q(T)_{\bigl| {\rm pert.th.}} &=& \frac{6}{\pi^2} + {\rm O}(g^3) \; , \\
R_{4,2}^Q(T)_{\bigl| {\rm pert.th.}} &=& \frac{34}{15 \pi^2} +
{\rm O}(g^3) \; . \label{R42qideal}
\end{eqnarray}
These values are also shown as solid lines in Fig.~\ref{fig:charge}.

\subsection{Fluctuations in the QGP}

The considerations presented in the previous section for hadrons
below $T_c$ carry over to any thermodynamic system containing
heavy quasi-particle states that are build up from quarks and
gluons. As long as their masses are significantly larger than
$T_c$, their contribution to the thermodynamics can be treated
within the Boltzmann approximation. States with mass $m_i$
containing a net number of $k_u$ up quarks and $k_d$ down quarks
will contribute to the pressure of the system with
\begin{equation}
\frac{p_i}{T^4} = F(T,m_i)
\cosh\left( k_u \mu_u +k_d \mu_d \right) \quad ,
\label{generic}
\end{equation}
where $F(T,m_i)$ is given by Eq.~(\ref{eq5}), i.e. it again
combines the contribution of a particle and its antiparticle.

A model containing heavy coloured bound states above $T_c$ has
been put forward by E. Shuryak and I. Zahed \cite{shuryak} to
explain the experimental evidence for strong interactions in the
hot and dense medium of quarks and gluons created in heavy-ion
collisions at RHIC. In the strongly coupled QGP model (sQGP), it
is assumed that, in addition to heavy quasi-particles (quarks and
gluons with large thermal masses), also a large number of coloured
bound states contribute to the thermodynamics in the
high-temperature phase of QCD. The model suggests, that aside from
states carrying net quark number $1$ and charge $Q=1/3$ and $2/3$
there will also be states with net quark number $2$ and charge
$Q=1$ and $4/3$ contributing to the thermodynamics in the QGP.
From the discussion given in the previous section it should be
obvious that this must have consequences for quark number and
charge fluctuations in the QGP and the ratios $R_{4,2}^q$ and
$R_{4,2}^Q$ should be sensitive to these new states. The lattice
results shown in Fig.~\ref{fig:charge}, can  thus put bounds on
their relative abundance.

For our discussion of baryon number and charge fluctuations, the
quark--quark and quark--gluon bound states are of particular
interest. The sQGP model suggests that the latter melt at $T_{\rm
melt}\equiv T_{qg}\simeq 2T_c$ while the former are only weakly
bound and melt already at $T_{\rm melt}\equiv T_{qq}\simeq 1.4
T_c$ \cite{shuryak,shuryak2}. It has been argued in
Ref.~\cite{shuryak} that the gradual disappearance of these states
could be taken into account by reducing their contribution to the
thermodynamic potential with increasing temperature. For the
corresponding weight factor, we will use here the parameterization
introduced in Ref.~\cite{shuryak}:
\begin{equation}
R_C(T,T_{melt}) = \frac{1}{1+\exp \left( C(T-T_{\rm melt})/T_c
\right)} \quad , \label{reduction}
\end{equation}
where $C$ controls the temperature interval over which the contribution
of a given bound state to the partition function gradually disappears.

In order to estimate the contribution of coloured states to the
fluctuations of baryon number and charge, we consider the
contribution of these states to the density-, (chemical
potential)-dependent part of the pressure,
\begin{eqnarray}
\frac{\Delta p}{T^4} &= & \frac{1}{2}\left( F_q(T) +R_2(T,2) F_{qg}(T) \right)
\left( \cosh(\mu_u/T) +\cosh(\mu_d/T) \right) \\
&&+\frac{1}{3} R_{C}(T,1.4) F_{qq}(T) ( \cosh(2 \mu_u/T) +
\cosh(2 \mu_d/T) \nonumber \\
&&+ \cosh((\mu_u+\mu_d)/T) ) \; , \nonumber \label{sQCDmu}
\end{eqnarray}
where we again  used the Boltzmann approximation as all states are
expected to have masses $m_i~ \gsim ~10 T_c$ \cite{shuryak}. We
will approximate the masses of quark--gluon and quark--quark bound
states as twice the quasi-particle mass of quarks, and  will thus
neglect mass shifts arising from a non-vanishing binding energy.
Moreover, we used $C=2$ for the quark--gluon bound states
\cite{shuryak} and will here only discuss the contribution of
quark--quark bound states as a function of $R_C$. With this we can
estimate the contribution of quark--quark states to the quadratic
and quartic quark number and charge fluctuations introduced in
Eq.~(\ref{fluctuations}):
\begin{eqnarray}
d_2^q &=& F_q(T) +\left( R_2(T,2) F_{qg}(T)+ 4 R_C (T,1.4)
F_{qq}(T)\right) \nonumber \\
d_4^q &=& F_q(T) +\left( R_2(T,2) F_{qg}(T)+ 16 R_C (T,1.4)
F_{qq}(T)\right) \; .\label{sQGPfluctq}
\end{eqnarray}
and
\begin{eqnarray}
d_2^Q &=& \frac{1}{9}\biggl( \frac{5}{2} F_q(T)+ \biggl(
\frac{5}{2} R_2(T,2) F_{qg}(T)+ \frac{11}{2}R_C (T,1.4) F_{qq}(T)
\biggr)\biggr) \nonumber \\
d_4^Q &=& \frac{1}{81}\biggl( \frac{17}{2} F_q(T)+ \biggl(
\frac{17}{2} R_2(T,2) F_{qg}(T)+ \frac{137}{2} R_C (T,1.4) F_{qq}(T)
\biggr)\biggr) \quad .
\label{sQGPfluctQ}
\end{eqnarray}
Apparently both ratios $R_{4,2}^{q,Q}$ are enhanced over the
corresponding ratios that would only contain states with the
quantum number of quarks\footnote{Note that this latter ratio
could also include a contribution from quark--gluon bound states.
Moreover, the observables discussed here are insensitive to the
presence of coloured quark--antiquark bound states, which could
exist in the QGP \cite{shuryak}.} owing to the presence of states
that carry two units of quark number. Assuming that the coloured
bound state masses are about twice as large as the quasi-particle
quark masses, we find at the melting temperature of the
quark--quark states $R_{4,2}^q = 1.55$ and $R_{4,2}^Q = 0.52$.
These values are shown as the dashed lines in
Fig.~\ref{fig:charge}. How quickly this enhancement disappears at
larger temperature depends on the precise choice of the
suppression factor $R_C$. At $T_{\rm melt}\simeq 1.4T_c$, the
contribution of quark--quark states should be suppressed below
10\% of that of an ordinary bound state ($R_C~\lsim ~0.1$) in
order to be compatible with the lattice results. This also could
mean that the melting temperature is substantially smaller
($T_{\rm melt}~\lsim ~1.1T_c$).

We also note that the perturbative values given in
Eq.~(\ref{R42qideal}) are obtained for a massless
quarks--antiquarks (fermion) gas. These ratios are smaller  by
about 50\% than the corresponding values in the Boltzmann
approximation. This indicates the uncertainties inherent in the
estimates given above and suggests that if coloured bound states
exist in the QGP and contribute to the thermodynamics, this
presumably has to be discussed in terms of a more complex
interaction pattern, which also takes into account quantum
effects.

\section{Conclusions}

We have discussed some generic features of quark number, isospin
and charge fluctuations in 2-flavour QCD. We argue that the ratio
of quartic and quadratic fluctuations are sensitive observables
that can directly provide information on the constituents of the
thermal medium that carry net quark number and electric charge,
respectively. We have shown that these ratios,  below the QCD
transition temperatures,  are in reasonable agreement with a
hadronic resonance gas. Above $T_c$ the ratios rapidly drop and
approach the high-temperature ideal-gas values. This suggests
that, already for $T~\gsim ~1.5T_c$ quark number and charge are
predominantly carried by states with the quantum number of quarks.
This leaves little room for the contribution of quark--quark bound
states. It is conceivable that these constraints  would be even
more stringent in 3-flavour QCD calculations, as the relevant
states in the sQGP model are proportional to $n_f^2$ while the
relevant degrees of freedom in the (perturbative) high-temperature
limit of QCD are only proportional to $n_f$. A similar conclusion
on the incompatibility of the sQGP model with lattice results on
hadronic fluctuations has been drawn recently in \cite{koch}.

It would, of course, be interesting to verify these results in
experimental studies of event-by-event fluctuations
\cite{all,probe} of baryon number or charge. However, if indeed
the medium created in a heavy-ion collision is strongly
interacting and  rapidly equilibrates there is little hope to find
evidence for fluctuations originate from the QGP. At best, one may
hope to study the quadratic and quartic fluctuations at
freeze-out. Also this, however, would be interesting, as the
observation of the importance of doubly charged hadrons leading to
$R_{4,2}^Q\simeq 1.5$ could give further support to the validity
of the resonance gas model.

\section*{Acknowledgements}
\addcontentsline{toc}{section}{Acknowledgments}

This work was partly supported by the  KBN under grant 2P03
(06925), the DFG under grant KA 1198/6-4, and the GSI
collaboration grant BI-KAR. The work of FK has been partly
supported by a contract DE-AC02-98CH1-886 with the U.S. Department
of Energy. K.R. thanks to E. Shuryak for interesting discussions.


\end{document}